\documentclass[10pt, conference]{IEEEtran}
\IEEEoverridecommandlockouts
\usepackage{cite}
\usepackage{amsmath,amssymb,amsfonts}
\usepackage{mathptmx}
\usepackage{algorithmic}
\usepackage{graphicx}
\usepackage{multirow}
\usepackage{textcomp}
\usepackage{xcolor}
\def\BibTeX{{\rm B\kern-.05em{\sc i\kern-.025em b}\kern-.08em
    T\kern-.1667em\lower.7ex\hbox{E}\kern-.125emX}}
\usepackage{blindtext}
\usepackage[caption=false]{subfig} 
\usepackage{pgfplots} % Create normal/logarithmic plots in two and three dimensions
\pgfplotsset{width=7cm,compat=1.14}
\usepgfplotslibrary{statistics}
\usepackage{tikz} % Create PostScript and PDF graphics
\usetikzlibrary{calc,patterns,shapes}
\usepackage{booktabs} %toprule, midrule, bottomrule, ..
\usepackage{listings}
\usepackage[breaklinks]{hyperref}

\newenvironment{customlegend}[1][]{%
	\begingroup
	\csname pgfplots@init@cleared@structures\endcsname
	\pgfplotsset{#1}%
}{%
	\csname pgfplots@createlegend\endcsname
	\endgroup
}%
\def\addlegendimage{\csname pgfplots@addlegendimage\endcsname}

\newcommand{\bq}{\begin{equation}}
\newcommand{\eq}{\end{equation}}
\newcommand{\bytes}{\mbox{bytes}}
\newcommand{\byte}{\mbox{byte}}
\newcommand{\second}{\mbox{s}}

\newcommand{\flop}{\mbox{flop}}

\newcommand{\cycle}{\mbox{cy}}

\newcommand{\cycles}{\mbox{cy}}

\newcommand{\BC}{\mbox{\byte/\cycle}}
\newcommand{\GBS}{\mbox{G\byte/\second}}

\newcommand{\GFS}{\mbox{G\flop/\second}}

\newcommand{\GHZ}{\mbox{GHz}}

\newcommand{\BF}{\mbox{\byte/\flop}}

\newcommand{\GB}{\mbox{GB}}
\newcommand{\KB}{\mbox{kB}}

\newcommand{\MiB}{\mbox{MiB}}
\newcommand{\KiB}{\mbox{KiB}}

\newcommand{\muops}{\mbox{$\mu$-ops}}

\newcommand{\ecmspace}{\,}

\newcommand{\epsep}{\rceil}
\newcommand{\ecmp}[4]{\mbox{$\left\{{#1}\ecmspace\epsep\ecmspace {#2}\ecmspace\epsep\ecmspace {#3}\right\}\ecmspace{#4}$}}
\newcommand{\ecme}[4]{\mbox{$\left({#1}\ecmspace\epsep\ecmspace {#2}\ecmspace\epsep\ecmspace {#3}\right)\ecmspace{#4}$}}
\newcommand{\sellcs}{SELL-$C$-$\sigma$}

\newcommand{\likwidperfctr}{\texttt{likwid-perfctr}}

\newcommand{\afx}{A64FX}
\newcommand{\spmv}{SpMV}
\newcommand{\mathspace}{\text{ }}

\definecolor{applegreen}{rgb}{0.55, 0.71, 0.0}

\begin{document}

\title{Performance Modeling of Streaming Kernels and Sparse Matrix-Vector Multiplication
  on \afx\\
%\title{An Execution-Cache-Memory Performance Model for the \afx\ CPU\\
\thanks{This work was supported in part by KONWIHR and by DFG in the framework of SFB/TRR 55.}
}

\author{\IEEEauthorblockN{Christie Alappat, Jan Laukemann, Thomas Gruber,\\Georg Hager, Gerhard Wellein}
\IEEEauthorblockA{\textit{Erlangen Regional Computing Center} \\
\textit{Friedrich-Alexander-Universit\"at Erlangen-N\"urnberg}\\
Erlangen, Germany\\
christie.alappat@fau.de}
\and
\IEEEauthorblockN{Nils Meyer, Tilo Wettig}
\IEEEauthorblockA{\textit{Department of Physics} \\
\textit{University of Regensburg}\\
Regensburg, Germany \\
tilo.wettig@ur.de}}
%\and
%\and
%\and
%\and

\maketitle

\begin{abstract}
  The \afx\ CPU powers the current \#1 supercomputer on the Top500
  list.  Although it is a traditional cache-based multicore processor,
  its peak performance and memory bandwidth rival accelerator devices.
  Generating efficient code for such a new architecture requires a good understanding of its
  performance features. Using these features, we construct the
  Execution-Cache-Memory (ECM) performance
  model for the \afx\ processor in the FX700 supercomputer and
  validate it using streaming loops. We also identify architectural peculiarities
  and derive optimization hints. Applying the ECM model to sparse
  matrix-vector multiplication (\spmv), we motivate why the CRS matrix
  storage format is inappropriate and how the \sellcs\ format with
  suitable code optimizations can achieve bandwidth
  saturation for \spmv.  
\end{abstract}


\section{Introduction}\label{sec:intro}

\subsection{Motivation: The \afx\ CPU}

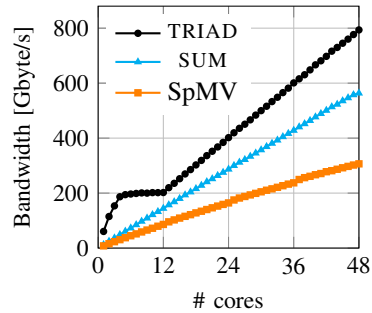
\begin{figure}[tb]
	\begin{minipage}{0.6\linewidth}
	%
	%  \tikzsetnextfilename{#2}%
	\begin{tikzpicture}%[y=.2cm, x=.7cm]%,font=\sffamily]
\centering
\begin{axis}[
%%scale only axis,
height=4.8cm,
width=.95\linewidth,
grid=both,
tick align=inside,
%%enlarge y limits={value=.1,upper},
%%axis x line*=bottom,
%%axis y line*=left,
%%y axis line style={opacity=0},
%tickwidth=0pt,
%enlarge x limits=true,
xlabel={\# cores},
ylabel={Bandwidth [\GBS]},
%%%%%%xlabel style={},
%%%%%%ylabel style={},
%%%%%%x tick label style={},
%%%%%%y tick label style={},
% minor x tick num = 7,
xmin=0,
xmax=48,
ymin=0,
ymax=880,
xtick = {0,12,24,36,48},
legend pos=north west,
legend style={draw=none},
label style={font=\small},
tick label style={font=\small},
]
\addplot[black, mark=*, mark size=1pt, %mark options={fill=white}, 
thick] table[x expr=\thisrowno{0}, y expr=\thisrowno{5}*0.001*4/3, header=true, col sep=comma] {tikz_data/bwbench_u1.dat};
\addlegendentry{\textsc{triad}}

\addplot[cyan, mark=triangle*, mark size=1pt, %mark options={fill=white}, 
thick] table[x expr=\thisrowno{0}, y expr=\thisrowno{2}*0.001, header=true, col sep=comma] {tikz_data/bwbench_u1.dat};
\addlegendentry{\textsc{sum}}

\addplot[orange, mark=square*, mark size=1pt, %mark options={fill=white},
thick] table[x expr=\thisrowno{0}, y expr=\thisrowno{1}*0.5*(12+28/26.8), header=true, col sep=comma] {tikz_data/scaling_hpcg_csr_u1.dat};
\addlegendentry{SpMV}

\end{axis}
%\node[fill=white] at (1.1,2.5) {(a) STREAM};
\end{tikzpicture}%

	\end{minipage}
	\begin{minipage}{0.38\linewidth}
			\caption{Scaling of STREAM \textsc{triad}, \textsc{sum} reduction, and \spmv\
			in CRS format with the HPCG matrix using gcc version 10.1.1. The working set size
			is 4\,\GB\ for \textsc{triad} and \textsc{sum};
			for HPCG the dimension is $128^3$.}
			\label{fig:csr_scaling}
	\end{minipage}
\vspace{-1em}
\end{figure}

The \afx\ CPU is used in parallel computer designs
from Fujitsu. It exists in several variants, the most basic
of which (used, e.g., in the Fujitsu FX700 system) comprises 48 cores
running at 1.8\,\GHZ\ (see Table~\ref{table:testbed} for fundamental
data). At a
%peak performance of 2.8\,\TFS\ and a theoretical memory
%bandwidth of 1.0\,\TBS, the
machine balance of 0.37\,\BF, the architecture is expected to deliver
a high fraction of its peak performance for optimized code. The chip
is divided into four groups of twelve cores (\emph{core memory groups}
[CMGs]), each of which accesses its own ccNUMA domain. The 64\,\KiB\ L1 cache
is core local, while 8\,\MiB\ of L2 are shared among the cores of each
CMG. Figure~\ref{fig:csr_scaling}
shows bandwidth scaling using compiler-generated OpenMP code
with compact pinning for three elementary operations:
the STREAM \textsc{triad} (\verb.a[i]=b[i]+s*c[i].), a sum reduction (\verb.s+=a[i].),
and a sparse matrix-vector multiplication with the HPCG~\cite{DongaraHPCG} matrix using
the Compressed Row Storage (CRS) format. All of these
should be strongly memory bound, but only \textsc{triad} shows the
typical saturation pattern within the first ccNUMA domain; the other two,
albeit scalable, top out at only 40\% and 70\% of the maximum
\textsc{triad} bandwidth, respectively. One goal of
our work is to investigate the reasons for this failure and how to
mitigate it. The Execution-Cache-Memory (ECM) performance model~\cite{Hager:2012,sthw15,JSFI310}
will be instrumental in this, leading to valuable insights into
performance bottlenecks of this new CPU architecture.
%\begin{figure}[tb]
%  \includegraphics*[width=\linewidth]{fx700_cpu}
%  \caption{Do we have space for this? \CAcomm{I dont think so}}
%\end{figure}

\subsection{Brief overview of the ECM model}

Full coverage of the ECM model is beyond the scope of this work. We give a
brief overview and refer to the most recent publication~\cite{JSFI310} for
details.

The model considers execution time contributions for steady-state loops
from the core (assuming all data is in L1), data paths in the cache hierarchy, and
the memory interface. For the core component, the loop's assembly code
is analyzed
%manually or using the OSACA tool~\cite{OSACA2019}
for predictions of
optimal throughput, critical path, and longest loop-carried dependency (the current development branch of the OSACA tool~\cite{OSACA2019} has preliminary support for \afx).
Data transfer volumes through the memory hierarchy are
obtained either by manual analysis or by the Kerncraft~\cite{Hammer:2017} tool; together with
the known bandwidths of all data paths, time contributions for
L1-L2 and L2-memory transfers are obtained. A machine model is constructed
that makes assumptions on how all these contributions overlap in
time in order to arrive at a runtime prediction for a given number
of loop iterations. For the benchmarks under investigation
here, manual predictions are straightforward and the Kerncraft tool
is not needed.
%The current version of OSACA already supports \afx.\GHcomm{correct?}

\subsection{Testbed and experimental methodology}
\begin{table}[!tb]
	\centering
	\caption{Key specifications of the \afx\ CPU in the FX700 system.}
	\label{table:testbed}
%	\resizebox{\textwidth}{!}{%
		\begin{tabular}{l c }
			\toprule
			Microarchitecture       & \afx  \\
			Supported core frequency& 1.8\,\GHZ   \\
			Cores/threads           & 48/48  \\        
			Instruction set         & Armv8.2-A+SVE \\
                        Max. SVE vector length  & 512 bit\\
                        Cache line size         & 256\,\bytes\\
			L1 cache capacity       & 48$\times$64\,\KiB  \\
			L1 bandwidth per core ($b_{\mathrm{Reg}\leftrightarrow\mathrm{L1}}$)	& 128\,B/cy LD $\oplus$ 64\,B/cy ST\\
			L2 cache capacity       & 4$\times$8\,\MiB \\
			L2 bandwidth per core ($b_{\mathrm{L1}\leftrightarrow\mathrm{L2}}$) 	&  64\,B/cy LD, 32\,B/cy ST \\
			Memory configuration    & 4$\times$HBM2 \\
			CMG theor. mem. bandwidth  & 230\,\GBS\,=\,128\,\BC \\
                        CMG \textsc{triad} bandwidth & 210\,\GBS\,=\,117\,\BC\\
                        CMG read-only bandwidth & 225\,\GBS\,=\,125\,\BC\\
			Page size & 64\,\KiB  \\
			L1 translation lookaside buffer & 16 entries  \\
			L2 translation lookaside buffer & 1024 entries\\
			%\midrule
			%        %Compiler               & Intel icc 19.0 update 2   \\
			%        %Optimization flags     & \texttt{-O3 -xHost}       \\
			%        %                       & \texttt{-mavx2 -mfma}    \\
			\bottomrule
		\end{tabular}
%	} % resizebox
\end{table}
%\stepcounter{footnote}\footnotetext{
%	The total L2 bandwidth per CMG is further capped at 512\,B/cy}

%An overview of the investigated Fujitsu FX700 system is provided in Table~\ref{table:testbed}.
All experiments were carried out on QPACE~4, the Fujitsu FX700 system running CentOS 8.2 at the physics department of the
University of Regensburg.
The core clock frequency is fixed to 1.8\,\GHZ\ on this machine.
%The core frequency of 1.8\GHZ is not adjustable and can be considered as fixed during computational load.
All code was compiled with gcc~10.1.1, using options \texttt{-Ofast} \texttt{-msve-vector-bits=512} \texttt{-march=armv8.2-a+sve}.
%to enable SVE SIMD instructions.
Apart from the cases shown in Fig.~\ref{fig:csr_scaling},
SVE vector intrinsics (ACLE~\cite{ACLE}) were employed to have better control over code generation. All benchmarks were run in double
precision, leading to a vector length (VL) of eight elements.
Performance event counting
%and loop microbenchmarking \CAcomm{not for microbenchmarking likwid-bench is not yet ready}
%for standard loop kernels
was done
with \likwidperfctr~\cite{Treibig:2010:2} v5.0.1, whose current development branch contains
support for \afx. For benchmarking individual machine instructions we employed the
ibench~\cite{ibench} framework and cross-checked our results with the \afx\
Microarchitecture Manual~\cite{a64fx_manual}.
We do not show any run-to-run statistics as the fluctuations in measurements were below 3\%.
Information about how to reproduce the results in this paper can be found in the artifact description~\cite{artif}.

This paper is organized as follows:  Section~\ref{sec:arch} provides an analysis of the
in-core architecture, including parts of the SVE instruction set, and
the memory hierarchy of the \afx. In Sect.~\ref{sec:ecm} we use microbenchmarks
to construct and validate the ECM performance model for serial and parallel steady-state
SVE loops. The insights gained are then used in Sect.~\ref{sec:spmvm} to revisit
the \spmv\ kernel and motivate why an appropriate matrix storage as well as
sufficient unrolling are required to achieve bandwidth saturation. Finally,
Sect.~\ref{sec:summary} gives a summary and an outlook to future work.

%\JLcomm{Did someone use the GCC 8 as well? Should we mention -O flags?} 
%FX700, gcc~8.3.1 or gcc~10.1.1, ibench, osaca

%\subsection{Contributions}  %% no space for that
%
%TODO -- probably fuse into 1.1

\section{Architectural analysis}\label{sec:arch}

\subsection{In-core}

\begin{table}[!tb]
	\centering
	\caption{In-core instruction throughput and latency (if applicable) for selected instruction forms.}
	\label{table:incore-instructions}
%	\resizebox{\textwidth}{!}{%
		\begin{tabular}{l c c}
			\toprule
			\multirow{2}{*}{\textbf{Instruction}} & \textbf{Reciprocal} & \multirow{2}{*}{\textbf{Latency [cy]}} \\
			& \textbf{Throughput [cy]} & \\
			\hline
			\texttt{ld1d} (standard)	& 0.5 & 11 \\
			\texttt{ld1d} (gather, simple stride)& 2.0 & $\geq$ 11 \\
			\texttt{ld1d} (gather, complex stride)& 4.0 & $\geq$ 11 \\
			simple gather + std. load			& 3.5 & -- \\ %\emph{N/A} \\
			complex gather + std. load			& 5.5 & -- \\ %\emph{N/A} \\
			\texttt{st1d} (standard)			& 1.0 & -- \\ %\emph{N/A} \\
			\texttt{fadd}                   & 0.5 & 9 \\
			\texttt{fmad}                   & 0.5 & 9 \\
			\texttt{fmla}                   & 0.5 & 9 \\
			\texttt{fmul}                   & 0.5 & 9 \\
			\texttt{fadda} (512bit) & 18.5 & 72 \\
			\texttt{faddv} (512bit) & 11.5 & 49 \\
			\texttt{while\{le|lo|ls|lt\}}                   & 1.0 & 1 \\
			\bottomrule
		\end{tabular}
%	} % resizebox
\end{table}

For creating an accurate in-core model of the \afx\ micro\-archi\-tecture, we analyze different \emph{instruction forms}~\cite{JLBA17}, i.e., assembly instructions in combination with their operand types, based on the methodology introduced in~\cite{OSACA2018}. %\footnote{The \afx\ portmodel is also available in OSACA for static in-core runtime analysis. See \href{https://github.com/RRZE-HPC/OSACA}{github.com/RRZE-HPC/OSACA}}
%We use ibench~\cite{ibench} for microbenchmarking throughput and latency and validate our results with the information obtained from the \afx~Microarchitecture~Manual~\cite{a64fx_manual}.
Table~\ref{table:incore-instructions} shows a list of instruction forms relevant for this work.
Standard SVE load (\texttt{ld1d}) instructions have a reciprocal throughput of 0.5\,\cycle, while stores (\texttt{st1d}) have 1\,\cycle.
%Furthermore, in contrast to current micro-architectures by other vendors,
The throughput of gather instructions depends on the distribution of addresses:
``Simple'' access patterns are stride 0~(no stride), 1~(consecutive load), and 2, while larger strides and irregular patterns are considered ``complex.'' The former have lower reciprocal throughput and latency than the latter.
However, when occurring in combination with a standard LD, we can observe an increase of reciprocal throughput by 1.5\,\cycles\
instead of the expected 0.5\,\cycle.
This is caused by the
%extension of the critical path due to the
dependency of the gather instruction on the preceding index load operation, which the out-of-order~(OoO) execution
cannot hide completely.

%While we cannot pinpoint the reason for this exactly, we assume this is due to an overhead in the address generation units when ``switching'' between the two LD modes.

Note also the rather long latencies for arithmetic operations such as MUL, ADD, and FMA compared to other state-of-the-art architectures (e.g., on Intel Skylake or AMD Zen2 these are between 3\,\cycles\
and 5\,\cycles).
Each core's front end has an instruction buffer with $6\times8$ entries, which can feed six instructions per cycle to the decoder. The decoder feeds up to four \muops\ per cycle to two pairs of reservation stations, which have ten (address generation and load units) or 20 (execution pipelines and store units) entries. 
These attributes emphasize the importance of a compiler that is capable of exploiting the theoretical in-core performance by intelligent code generation.
The small reservation stations in combination with high instruction latencies can result in inefficient OoO execution.
While these constraints cannot be overcome completely, appropriate loop unrolling, consecutive addressing,
and interleaving of different instruction types have shown to be beneficial in our benchmarks;
see Sections~\ref{sec:ecm} and \ref{sec:spmvm} for details.

The SVE instruction set introduced a ``\texttt{while}\{cond\}'' instruction to set predicate registers
%according to the elements in vector registers in a length-agnostic way and, therefore,
in order to eliminate remainder loops (see Sect.~\ref{sec:spmvm} for details).
A port conflict analysis revealed that this instruction does not collide with floating-point instructions or data transfers, so it does not impact the kernel runtime compared to non-SVE execution.
%Hence, the additional operation compared to non-SVE code does not effect the investigated kernels in a negative way.

%\JLcomm {Here the OoO LOAD/ADD/STORE histogram can be added, but probably there is no space}

%\begin{itemize}
%\item Instruction latency and throughput
%\begin{itemize}
%    \item sub-optimal OoO execution due to 8$\times$6 entries instruction buffer, 10--20 entries reservation stations
%    \item latencies/loop-carried dependencies detectable and modelable
%\end{itemize}
%\item Guidelines for efficient in-core code
%\begin{itemize}
%    \item streaming STOREs
%    \item unrolling to overcame latencies
%\end{itemize}
%\end{itemize}

\subsection{Memory hierarchy}
%Each A64FX core makes use of a private L1~data~cache and a L2-cache shared between the cores in one CMG, as stated in Table~\ref{table:testbed}.
While parallel load/store from and to L1D is possible for general-purpose and NEON registers, different types of SVE data transfer instructions in L1D cannot be executed in one cycle:
Using SVE, one \afx\ core can \emph{either} load up to $2\times64\,\BC$ \emph{or} store $64\,\BC$ from/to L1.
%This results in a theoretical L1 cache bandwidth of $230\,\GBS$ for LOADs and $115\,\GBS$ for STOREs.
The L2 cache can deliver $64\,\BC$ to one L1D but tops out at $512\,\BC$ per CMG.
The L1D-L2 write bandwidth is half the load bandwidth, i.e., $32\,\BC$ per core, and is capped at $256\,\BC$ per CMG.
%The theoretical L1D--L2 bandwidth per CMG therefore is $58\,\GBS$ and $115\,\GBS$ for data transfer into L1D and into L2, respectively.
Finally, the maximum bandwidth between memory and L2 cache is $128\,\BC$ per CMG for loading and $64\,\BC$ per CMG for storing data, resulting in a theoretical total peak load bandwidth of $922\,\GBS$~($230\,\GBS$ per CMG) and peak store bandwidth of $461\,\GBS$~($115\,\GBS$ per CMG). In practice, about 91\% of the peak load bandwidth can be attained using
the STREAM \textsc{triad} benchmark and 98\% using a read-only benchmark (see Table~\ref{table:testbed}).
These measured bandwidths will be used as baselines for the memory transfer bandwidth in the ECM model.
%For modeling read-only kernels, we will use the measured read-only bandwidth of 225\,\GBS\ per CMG.

\section{Construction of the ECM model}\label{sec:ecm}
\label{sec:ECM}
\subsection{Overlap hypothesis}
\label{subsec:overlap_hypo}
In order to find out which of the time contributions for data
transfers through the cache hierarchy overlap, measurements for a test
kernel are compared with predictions based on different hypotheses;
see~\cite{JSFI310} for an in-depth description of the process.
If a hypothesis works for the test kernel, it is tested against
a collection of other kernels with different characteristics.

\begin{figure}[tb]
	\centering
	%
	%  \tikzsetnextfilename{#2}%
	\begin{tikzpicture}%[y=.2cm, x=.7cm]%,font=\sffamily]
    \centering
    \begin{axis}[
            %%scale only axis,
		height=4.5cm,
		width=0.95*\linewidth,
		xmajorgrids,
		ymajorgrids,
        tick align=inside,       
        %%enlarge y limits={value=.1,upper},
        %%axis x line*=bottom,
        %%axis y line*=left,
        %%y axis line style={opacity=0},
        %tickwidth=0pt,
        %enlarge x limits=true,
        xlabel={Size (\KB)},
        ylabel={Runtime (cy/VL)},
        axis y line*=left,
        xmode=log,
        %ylabel={Effective BW [GB/s]},
        %%%%%%xlabel style={},
        %%%%%%ylabel style={},
        %%%%%%x tick label style={},
        %%%%%%y tick label style={},
        %minor x tick num = 7,
        xmin=4,
        xmax=2000000,
        ymin=0,
        ymax=9,
        %ymax=5,
        %xtick = {0,2500,5000,7500,10000},
        %xticklabels = {0,2500,5000,7500,10000},
        %ytick = {0,1,2,3,4,5},
        %x tick label style={font={\small}},
        %x tick label style={rotate=30, anchor=north east, inner sep=0mm, font={\normalsize}},
        %ytick = {0, 25, 50, 75, 100, 125, 150},
        %ytick = {0,20,40,60,80,100,120,140},
        legend pos=south east,
        legend style={draw=none},
        label style={font=\small},
        tick label style={font=\small},
    ]
        \addplot[mark=none, black, thick, mark size=0.75pt, line width=1] table[x expr=\thisrowno{1}, y expr=\thisrowno{2}, header=true, col sep=comma] {tikz_data/stream/u1.txt}; \label{stream_unroll_one}
        %\addlegendentry{u=1}
        \addplot[mark=none, applegreen,  thick, mark size=0.75pt, line width=1] table[x expr=\thisrowno{1}, y expr=\thisrowno{2}, header=true, col sep=comma] {tikz_data/stream_unaligned/u8.txt}; \label{stream_unaligned}
	%	\addlegendentry{u=8, malloc} 
       %\addplot[brown, mark options={fill=white}, thick] table[x expr=\thisrowno{1}, y expr=\thisrowno{2}, header=true, col sep=comma] {tikz_data/stream/u2.txt};
       % \addlegendentry{u=2}
        \addplot[mark=none, cyan,  thick, mark size=0.75pt, line width=1] table[x expr=\thisrowno{1}, y expr=\thisrowno{2}, header=true, col sep=comma] {tikz_data/stream/u8.txt};\label{stream_unroll_eight}
        %\addlegendentry{u=8}
     
        \addplot[red, dashed, very thick] table[x expr=\thisrowno{0}, y expr=\thisrowno{1}, header=true, col sep=comma] {tikz_data/ecm_stream.dat};\label{stream_ecm}
       % \addlegendentry{ECM}

        %\addlegendentry{Ideal}

    \end{axis}
	    \begin{axis}[
	%%scale only axis,
	height=4.5cm,
	width=0.95*\linewidth,
	tick align=inside,
	%%enlarge y limits={value=.1,upper},
	%%axis x line*=bottom,
	%%axis y line*=left,
	%%y axis line style={opacity=0},
	%tickwidth=0pt,
	%enlarge x limits=true,
	xlabel={Size (\KB)},
	ylabel={TLB misses/64 \KiB},
	axis y line*=right,
	xmode=log,
	%ylabel={Effective BW [GB/s]},
	%%%%%%xlabel style={},
	%%%%%%ylabel style={},
	%%%%%%x tick label style={},
	%%%%%%y tick label style={},
	minor y tick num = 0,
	xmin=4,
    xmax=2000000,
	ymin=0,
	ymax = 1.666,
	%ymax=5,
	%xtick = {0,2500,5000,7500,10000},
	%xticklabels = {0,2500,5000,7500,10000},
	%ytick = {0,1,2,3,4,5},
	%x tick label style={font={\small}},
	%x tick label style={rotate=30, anchor=north east, inner sep=0mm, font={\normalsize}},
	ytick = {0, 1, 2},
	legend pos=south east,
	legend style={draw=none},
	legend columns=1,
	label style={font=\small},
	tick label style={font=\small},
	]
	
%	\addlegendimage{/pgfplots/refstyle=stream_unroll_one}\addlegendentry{u=1}
%	\addlegendimage{/pgfplots/refstyle=stream_unaligned}\addlegendentry{u=8, malloc}
%	\addlegendimage{/pgfplots/refstyle=stream_unroll_eight}\addlegendentry{u=8}
%	\addlegendimage{/pgfplots/refstyle=stream_ecm}\addlegendentry{ECM}
		
	\addplot[mark=*, orange, thick, mark size=1pt, line width=0.5] table[x expr=\thisrowno{1}, y expr=\thisrowno{6}*1e-3*(64*1024)/24, header=true, col sep=comma] {tikz_data/stream_w_tlb/u8.txt};
	%\addlegendentry{L2 TLB}

	%\addlegendentry{Ideal}
	
	\end{axis}
\node[fill=white] at (0.88,2.6) {(a) \textsc{triad}};
\end{tikzpicture}

\vspace{0.5em}
\begin{tikzpicture}%[y=.2cm, x=.7cm]%,font=\sffamily]
\centering
\begin{axis}[
%%scale only axis,
height=4.5cm,
width=0.56*\linewidth,
xmajorgrids,
ymajorgrids,
tick align=inside,
%%enlarge y limits={value=.1,upper},
%%axis x line*=bottom,
%%axis y line*=left,
%%y axis line style={opacity=0},
%tickwidth=0pt,
%enlarge x limits=true,
xlabel={Size (\KB)},
ylabel={Runtime (cy/VL)},
xmode=log,
%ylabel={Effective BW [GB/s]},
%%%%%%xlabel style={},
%%%%%%ylabel style={},
%%%%%%x tick label style={},
%%%%%%y tick label style={},
%minor x tick num log= 1,
xmin=100,
xmax=400000,
ymin=0,
ymax=20,
%  ymax=9,
%ymax=5,
%xtick = {0,2500,5000,7500,10000},
%xticklabels = {0,2500,5000,7500,10000},
%ytick = {0,1,2,3,4,5},
%x tick label style={font={\small}},
%x tick label style={rotate=30, anchor=north east, inner sep=0mm, font={\normalsize}},
%ytick = {0, 25, 50, 75, 100, 125, 150},
%ytick = {0,20,40,60,80,100,120,140},
xtick={10,100,1000,10000, 100000, 1000000},
legend pos=south east,
legend style={draw=none},
label style={font=\small},
tick label style={font=\small},
]
\addplot[mark=*, black, thick, mark size=0.75pt, line width=0.5] table[x expr=\thisrowno{3}, y expr=\thisrowno{4}, header=true, col sep=comma] {tikz_data/stencil/u1.txt}; \label{stencil_unroll_one}
%\addlegendentry{u=1}
%	\addlegendentry{u=8, malloc} 
%\addplot[brown, mark options={fill=white}, thick] table[x expr=\thisrowno{1}, y expr=\thisrowno{2}, header=true, col sep=comma] {tikz_data/stream/u2.txt};
% \addlegendentry{u=2}
\addplot[mark=square*, cyan,  thick, mark size=0.75pt, line width=0.5] table[x expr=\thisrowno{3}, y expr=\thisrowno{4}, header=true, col sep=comma] {tikz_data/stencil/u8.txt};\label{stencil_unroll_eight}
%\addlegendentry{u=8}

\addplot[red, dashed, very thick] table[x expr=\thisrowno{0}, y expr=\thisrowno{1}, header=true, col sep=comma] {tikz_data/ecm_stencil.dat};\label{stencil_ecm}

%\addlegendentry{Ideal}

\end{axis}
\node[fill=white] at (0.9,2.6) {(b) \textsc{2d5pt}};
\end{tikzpicture}
\hspace{-0.5em}
\begin{tikzpicture}%[y=.2cm, x=.7cm]%,font=\sffamily]
\centering
\begin{axis}[
%%scale only axis,
height=4.5cm,
width=0.56*\linewidth,
xmajorgrids,
ymajorgrids,
tick align=inside,
%%enlarge y limits={value=.1,upper},
%%axis x line*=bottom,
%%axis y line*=left,
%%y axis line style={opacity=0},
%tickwidth=0pt,
%enlarge x limits=true,
xlabel={Size (\KB)},
xmode=log,
%ylabel={Effective BW [GB/s]},
%%%%%%xlabel style={},
%%%%%%ylabel style={},
%%%%%%x tick label style={},
%%%%%%y tick label style={},
minor y tick num = 1,
xmin=10,
xmax=700000,
ymin=0,
ymax=12,
%  ymax=9,
%ymax=5,
%xtick = {0,2500,5000,7500,10000},
%xticklabels = {0,2500,5000,7500,10000},
%ytick = {0,1,2,3,4,5},
%x tick label style={font={\small}},
%x tick label style={rotate=30, anchor=north east, inner sep=0mm, font={\normalsize}},
%ytick = {0, 25, 50, 75, 100, 125, 150},
%ytick = {0,20,40,60,80,100,120,140},
xtick={10,100,1000,10000, 100000, 1000000},
legend pos=south east,
legend style={draw=none},
label style={font=\small},
tick label style={font=\small},
]
\addplot[mark=*, black, thick, mark size=0.75pt, line width=0.5] table[x expr=\thisrowno{1}, y expr=\thisrowno{2}, header=true, col sep=comma] {tikz_data/sum/u1.txt}; \label{sum_unroll_one}
%\addlegendentry{u=1}
%	\addlegendentry{u=8, malloc} 
%\addplot[brown, mark options={fill=white}, thick] table[x expr=\thisrowno{1}, y expr=\thisrowno{2}, header=true, col sep=comma] {tikz_data/stream/u2.txt};
% \addlegendentry{u=2}
\addplot[mark=square*, cyan,  thick, mark size=0.75pt, line width=0.5] table[x expr=\thisrowno{1}, y expr=\thisrowno{2}, header=true, col sep=comma] {tikz_data/sum/u8.txt};\label{sum_unroll_eight}
%\addlegendentry{u=8}

\addplot[red, dashed, very thick] table[x expr=\thisrowno{0}, y expr=\thisrowno{1}, header=true, col sep=comma] {tikz_data/ecm_sum.dat};\label{sum_ecm}
% \addlegendentry{ECM}

%\addlegendentry{Ideal}

\end{axis}
\node[fill=white] at (0.9,2.6) {(c) \textsc{sum}};
\end{tikzpicture}%

	\resizebox{\linewidth}{!}{%
	%
	%  \tikzsetnextfilename{#2}%
	\begin{tikzpicture}
  \centering
  \begin{customlegend}[
      legend columns=5,
      legend style={
      draw=none,
      font=\small,
      anchor=north,
      /tikz/every even column/.append style={column sep=0.25cm}},
      legend entries={u=1, u=8, u=8+malloc, ECM, L2 TLB},% phenomenological ECM},
      ]
    \addlegendimage{mark=none, black, thick, mark size=1pt, line width=1}
    \addlegendimage{mark=none, cyan,  thick, mark size=1pt, line width=1}
    \addlegendimage{mark=none, applegreen,  thick, mark size=1pt, line width=1}
    \addlegendimage{red, dashed, very thick}
    \addlegendimage{mark=*, orange, thick, mark size=1pt, line width=1}
  \end{customlegend}
\end{tikzpicture}%

	}
	\vspace{-1em}
	\caption{Runtime of SVE loop kernels vs.\ problem size,
          comparing no unrolling (black) and eight-way unrolling (blue).
          Arrays were aligned to 1024-\byte\ boundaries.
          (a) STREAM \textsc{triad}; the extra green line denotes no special
          alignment (plain \texttt{malloc()}), while the orange data set
          denotes TLB misses per OS page, (b) 2d 5-point stencil, (c) sum reduction.
          For \textsc{2d5pt} the outer and inner dimension was set at a ratio of 1:2. 
        }
	\label{fig:stream_ecm}
\end{figure}
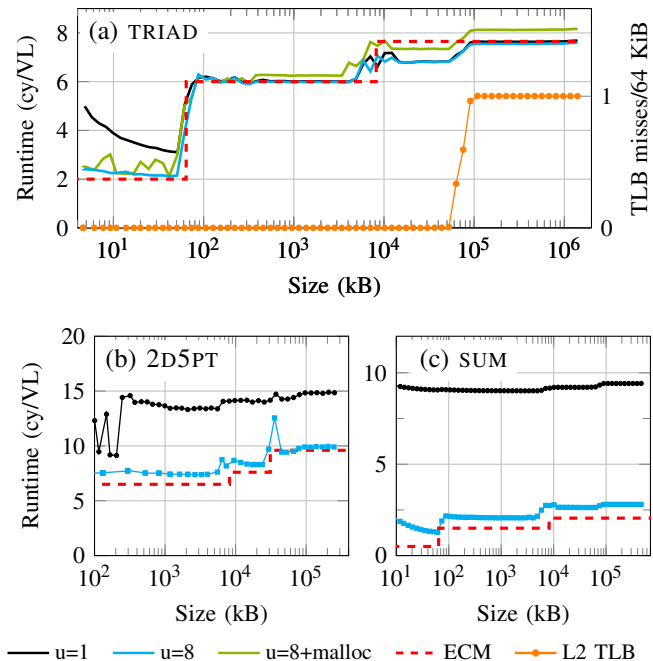
Here we use the STREAM triad kernel (\verb.a[i]=b[i]+s*c[i].) to
narrow down the possible overlap scenarios. This kernel has two LD,
one ST, and one FMA instruction per SVE-vectorized iteration.
Figure~\ref{fig:stream_ecm}a shows performance in cycles per
VL (i.e., eight iterations) for different code
variants: ``u=1'' denotes no unrolling (apart from SVE), and ``u=8''
is eight-way unrolled on top of SVE\@. Some level of manual unrolling
(typically eight-way) is always required for best in-core performance.
This is even more important in kernels where dependencies cannot
be resolved easily by the out-of-order logic.
In Fig.~\ref{fig:stream_ecm}b
we show data for a 2d five-point stencil, where SVE alone (without
further unrolling) is up to 2$\times$ slower than the eight-way unrolled
code, despite the lack of loop-carried dependencies. Figure~\ref{fig:stream_ecm}c shows data for a sum reduction,
which requires eight-way modulo variable expansion (MVE) on top
of SVE to achieve optimal performance due to the large latency
of the floating-point ADD instruction.

%\GHcomm{OK this way? we should
%swap c and b if we keep them} 
%\CAcomm{This is even more important in kernels which have
%	inherent dependencies like \textsc{sum} benchmarks dependency on ADD (see Figure~\ref{fig:stream_ecm}b)
%	and for complicated kernels where dependencies cannot
%	be resolved easily by the out-of-order logic as in the 
%	case of 2d5pt stencil (see Figure~\ref{fig:stream_ecm}c)}

Optimal performance beyond the L1 cache is only achieved when aligning all arrays
to 32-\byte\ boundaries; within L1, 512-\byte\ alignment is needed.
In our experiments we used 1024-\byte\ aligned arrays throughout.
The standard \verb.malloc(). function only guarantees
16-\byte\ alignment and leads to performance loss (``u=8+malloc'').

At a working set size of 64\,\MiB, performance drops significantly for all
three kernels.
We can correlate this with a sudden rise in L2 TLB misses (see
Fig.~\ref{fig:stream_ecm}a). Beyond this threshold,
exactly one TLB miss per 64\,\KiB\ page occurs.
%This high cost is unusual for
%a pure streaming code and
This leads to the conclusion that L2 TLB misses
are rather expensive on this machine.
%Whether this is a genuine
%architectural property or a misconfiguration of the benchmarking
%system is unknown at the time of writing.\GHcomm{too blunt?} 
A larger OS page size or other configuration changes might help improve the situation but are
left to future work. The effect will be ignored in the following.

\begin{figure}[tb]
  \includegraphics*[width=\linewidth]{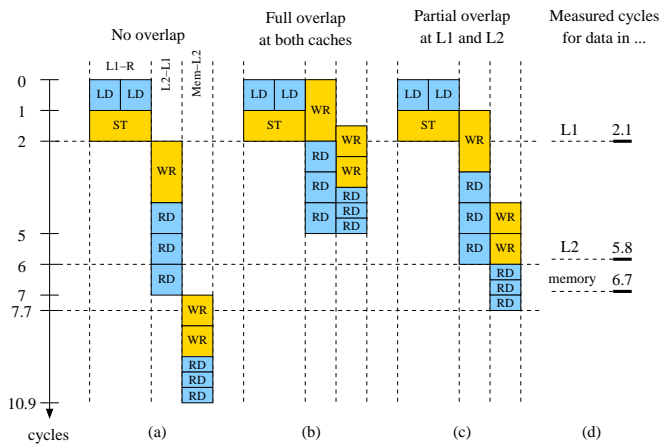}
  \caption{Comparing different overlap scenarios (a), (b), and (c)
    for data transfers in the memory hierarchy with measured cycles
    per VL (d) on the STREAM \textsc{triad} kernel.\label{fig:triad_ovl}}
\end{figure}

In order to arrive at an overlap hypothesis, we ignore the FMA
because it is fully overlapping when perfect
OoO processing is assumed. Figure~\ref{fig:triad_ovl}
compares three scenarios ((a), (b), and (c)) with measured cycles per
VL (d). Note that there is a large number
of possible overlap hypotheses, and we can only show a few here.
The one leading to the best match to the STREAM \textsc{triad} data is the following:
\begin{itemize}
\item L1D is partially overlapping:
  Cycles in which STs are retired in the core can overlap
  with L1-L2 (or L2-L1) transfers, but cycles with LDs retiring
  cannot.
  %Compute instructions overlap unconditionally with L1-L2
  %transfers.
\item L2 is partially overlapping:
  Cycles in which the memory interface writes data out to memory
  can overlap with transfers between L2 and L1, while memory read cycles
  cannot. 
\end{itemize}
As usual, the time required for compute instructions or, more generally,
non-data-transfer time in the core, fully overlaps with all data transfers.
Note that we include write-allocate transfers (due to store misses) in the
analysis.

%\begin{figure}[tb]
%  \includegraphics*[width=\linewidth]{triad_model}
%  \caption{We need dis too.}
%\end{figure}

\subsection{Validation of the single-core model for streaming kernels}
\label{subsec:validation}
With the in-core and data transfer models in place we can now test
the ECM model against a variety of loop kernels. Table~\ref{tab:validate}
shows a comparison of predictions and measurements. For each kernel,
three numbers represent the cycles per VL with the data set
in L1, L2, and memory, respectively. In case of the 2d five-point
stencil, three cases are shown: layer condition~(LC) satisfied at L1,
broken at L1, and broken at L2 (see~\cite{sthw15} for a comprehensive
coverage of layer conditions in the context of the ECM model).

The results have been obtained by running each kernel with unrolling factors
from 1 to 16 and taking the best result. Entries in red color have a
deviation from the model of 15\% or more. The strongest deviations
occur in L1: Even with eight-way MVE, the sum reduction cannot
achieve the architectural limit of 0.5\,\cycles/VL. A similar deviation
can be observed for the stencil kernels. We attribute this failure
to insufficient OoO resources: A modified  stencil
code without intra-iteration register dependencies achieves a performance
within 10\% of the prediction.
%which shows that the OoO engine is
%not able to hide the latencies by overlapping successive loop
%iterations if the critical path is the bottleneck.

Deviations from the model with L2 and memory working sets occur mainly
with kernels that have a single data stream. Indeed, the memory hierarchy
seems to work more effectively with multi-stream kernels, and a slight overlap
for memory reads can be observed for all of them. This could
be corrected by a refinement of the model if superior accuracy is
required.

%\begin{itemize}
%	\item Comment on accuracy of prediction L1-41\%.	L2=11\%,  Mem-12.67\%. Most of the deviation in 
%	L1 was for stencil and reduction kernels. 
%	Blame OoO, comment that for example stencil  
%	without the actual intra-dependencies reach 3.8\,\cycles\ in L1 (10\% to the model).
%	\item Comment that most deviation in L2 and Mem was when there was only a single stream
%	\item Comment a slight overlap in L2 and MEM
%	can be observed for kernels having higher than 1 stream. But the effect is on an average less than 5\%
%	and in actual practice this can be improved by adding 
%	a negative penalty, but we leave it out for brevity. 
%\end{itemize}

%\begin{figure}[tb]
%	\centering
%	\inputtikz{tikz}{ecm_stencil_sum}
	%\inputtikz{tikz}{ecm_stream_legend}
%	\vspace{-0.5em}
%	\caption{ECM of stencil and sum.} %with RCM
%	\label{fig:stencil_sum_ecm}
%\end{figure}

%\begin{itemize}
%\item STREAM triad / daxpy / STREAM copy (?) / init / load / 2d5pt \item Table %like in BBP-ECM paper
%\end{itemize}

\begin{table}
	\centering
	\caption{ECM model prediction and measurements in [cy/VL] for different streaming and
		stencil kernels. Red color indicates a deviation from the model of at least 15\%. The selected unrolling factor for each measurement is shown as a subscript.\label{tab:validate}}
\resizebox{\linewidth}{!}{%
	\begin{tabular}{lll}
		\toprule
		\textbf{Kernel}  &  \textbf{Predictions} & \textbf{Measurements} \\[0.5mm]
		\midrule
\textsc{copy} (\texttt{a[i]=b[i]}) & \ecmp{1.5}{4.5}{5.6}{} &  \ecme{1.6_{15}}{4.4_{13}}{5.1_{9}}{}\\[0.5mm]
\textsc{daxpy} (\texttt{y[i]=a[i]*x+y[i]})  &\ecmp{2.0}{5.0}{6.1}{} &  \ecme{2.1_{8}\mathspace}{4.7_{16}}{5.5_{6}}{}\\[0.5mm]
\textsc{dot}  (\texttt{sum+=a[i]*b[i]}) & \ecmp{1.0}{3.0}{4.1}{} &  \ecme{\textcolor{red}{1.7}_{8}\mathspace}{3.2_{5}\mathspace}{4.0_{5}}{}\\[0.5mm]
\textsc{init}  (\texttt{a[i]=s}) & \ecmp{1.0}{3.0}{3.5}{} &  \ecme{1.0_{13}}{\textcolor{red}{3.9}_{4}\mathspace}{\textcolor{red}{4.3}_{4}}{}\\[0.5mm]
\textsc{load}  (\texttt{load(a[i])}) & \ecmp{0.5}{1.5}{2.0}{} &  \ecme{\textcolor{red}{0.7}_{10}}{\textcolor{red}{1.9}_{15}}{\textcolor{red}{2.6}_{4}}{}\\[0.5mm]
\textsc{triad}  (\texttt{a[i]=b[i]+s*c[i]}) & \ecmp{2.0}{6.0}{7.7}{} &  \ecme{2.1_{8}\mathspace}{5.8_{12}}{6.7_{11}}{}\\[0.5mm]
\textsc{sum}  (\texttt{sum+=a[i]}) & \ecmp{0.5}{1.5}{2.0}{} &  \ecme{\textcolor{red}{1.1}_{11}}{\textcolor{red}{2.0}_{15}}{\textcolor{red}{2.5}_{13}}{}\\[0.5mm]
\textsc{sch\"{o}nauer}  (\texttt{a[i]=b[i]+c[i]*d[i]}) & \ecmp{2.5}{7.5}{9.7}{} &  \ecme{2.7_{7}\mathspace}{7.2_{10}}{8.4_{10}}{}\\[0.5mm]
\textsc{2d5pt} - LC satisfied  &  \ecmp{3.5}{6.5}{7.6}{} &  \ecme{\textcolor{red}{5.8}_{10}}{7.0_{10}}{8.0_{10}}{}\\[0.5mm]
\textsc{2d5pt} - LC violated in L1  &  \ecmp{3.5}{8.5}{9.6}{} &  \ecme{\textcolor{red}{5.8}_{10}}{8.6_{7}\mathspace}{9.4_{8}}{}\\[0.5mm]
\textsc{2d5pt} - LC violated  &  \ecmp{3.5}{8.5}{10.7}{} &  \ecme{\textcolor{red}{5.8}_{10}}{8.6_{7}\mathspace}{10.0_{7}}{}\\[0.5mm]
		\bottomrule
	\end{tabular}
}
\end{table}

 % deviation in percentage (meas-ecm)*100/ecm %
% Kernel, L1_dev, L2_dev, Mem_dev
% copy,   						6.67,  -2.22,  -8.93
% daxpy,   						5.00,  -6.00,  -9.84
% dot,  						70.00,   6.67,  -2.44
% init,   						0.00,  30.00,  22.86
% load,  						40.00,  26.67,  30.00
% stream,   					5.00,  -3.33, -12.99
% sum, 							120.00,  33.33,  25.00
% triad,   						8.00,  -4.00, -13.40
% 2d5pt - LC satisfied, 		65.71, 7.69, 5.26
% 2d5pt - LC violated in L1, 	65.71, 1.17, -2.08
% 2d5pt - LC violated, 			65.71, 1.17, -6.54
% total avg. 						41,		11,  12.67
% total wo stencil avg., 			32,		14,  15.68	

 % deviation in percentage (meas-ecm)*100/ecm with T_p=-0.5*T_mem for 2 or more streams%	
% Kernel, L1_dev, L2_dev, Mem_dev
% copy,   						6.67,  -2.22,   2.00
% daxpy,   						5.00,  -6.00,   0.00
% dot,  						70.00,   6.67,  14.29
% init,   						0.00,  30.00,  22.86
% load,  						40.00,  26.67,  30.00
% stream,   					5.00,  -3.33,  -1.47
% sum, 							120.00,  33.33,  25.00
% triad,   						 8.00,  -4.00,  -2.33
% 2d5pt - LC satisfied, 		65.71, 7.69, 13.47
% 2d5pt - LC violated in L1, 	65.71, 1.17, 3.87
% 2d5pt - LC violated, 			65.71, 1.17, 0.04
% total avg. 						41,		11,  10.48
% total wo stencil avg., 			32,		14,  12.24

\subsection{Multicore saturation}
\begin{figure}[tb]
	%
	%  \tikzsetnextfilename{#2}%
	\begin{tikzpicture}%[y=.2cm, x=.7cm]%,font=\sffamily]
\centering
\begin{axis}[
%%scale only axis,
height=4.5cm,
width=.42\linewidth,
grid=both,
tick align=inside,
%%enlarge y limits={value=.1,upper},
%%axis x line*=bottom,
%%axis y line*=left,
%%y axis line style={opacity=0},
%tickwidth=0pt,
%enlarge x limits=true,
xlabel={\# cores},
ylabel={Bandwidth [\GBS]},
%ylabel={Effective BW [GB/s]},
%%%%%%xlabel style={},
%%%%%%ylabel style={},
%%%%%%x tick label style={},
%%%%%%y tick label style={},
%minor x tick num = 7,
xmin=0,
xmax=12,
ymin=0,
ymax=295,
xtick = {0,4,8,12},
ytick = {0,  50, 100, 150, 200,250},
legend pos=south east,
legend style={draw=none},
label style={font=\small},
tick label style={font=\small},
]
\addplot[black, mark=*, mark size=1.5pt, %mark options={fill=white},
thick] table[x expr=\thisrowno{0}, y expr=\thisrowno{5}*0.001*4/3, header=true, col sep=comma] {tikz_data/bwbench_u8.dat};
\addplot[gray, mark=triangle*, mark size=1.5pt, %mark options={fill=white},
thick] table[x expr=\thisrowno{0}, y expr=\thisrowno{5}*0.001*4/3, header=true, col sep=comma] {tikz_data/bwbench_u1.dat};
%\addlegendentry{Measured}
\addplot[red, dashed, very thick] table[x expr=\thisrowno{0}, y expr=\thisrowno{4}, header=true, col sep=comma] {tikz_data/ecm_multicore.dat};
%\addlegendentry{ECM}

\end{axis}
\node[fill=white] at (0.88,2.6) {(a) \textsc{triad}};
\end{tikzpicture}
\hspace{-0.86em}
\begin{tikzpicture}%[y=.2cm, x=.7cm]%,font=\sffamily]
\centering
\begin{axis}[
%%scale only axis,
height=4.5cm,
width=.42\linewidth,
grid=both,
tick align=inside,
%%enlarge y limits={value=.1,upper},
%%axis x line*=bottom,
%%axis y line*=left,
%%y axis line style={opacity=0},
%tickwidth=0pt,
%enlarge x limits=true,
xlabel={\# cores},
%ylabel={Effective Bandwidth [GB/s]},
%ylabel={Bandwidth [\GBS]},
%ylabel={Effective BW [GB/s]},
%%%%%%xlabel style={},
%%%%%%ylabel style={},
%%%%%%x tick label style={},
%%%%%%y tick label style={},
%minor x tick num = 7,
xmin=0,
xmax=12,
ymin=0,
ymax=295,
xtick = {0,4,8,12},
ytick = {0,  50, 100, 150, 200, 250},
yticklabels={},
legend pos=south east,
legend style={draw=none},
label style={font=\small},
tick label style={font=\small},
]
\addplot[black, mark=*, mark size=1.5pt, %mark options={fill=white},
thick] table[x expr=\thisrowno{0}, y expr=\thisrowno{2}*0.001, header=true, col sep=comma] {tikz_data/bwbench_u8.dat};
\addplot[gray, mark=triangle*, mark size=1.5pt, %mark options={fill=white},
thick] table[x expr=\thisrowno{0}, y expr=\thisrowno{2}*0.001, header=true, col sep=comma] {tikz_data/bwbench_u1.dat};
%\addlegendentry{Measured}

\addplot[red, dashed, very thick] table[x expr=\thisrowno{0}, y expr=\thisrowno{5}, header=true, col sep=comma] {tikz_data/ecm_multicore.dat};
%\addlegendentry{ECM}

\end{axis}
\node[fill=white] at (0.88,2.6) {(b) \textsc{sum}};
\end{tikzpicture}
\hspace{-0.86em}
\begin{tikzpicture}%[y=.2cm, x=.7cm]%,font=\sffamily]
\centering
\begin{axis}[
%%scale only axis,
height=4.5cm,
width=.42\linewidth,
grid=both,
tick align=inside,
%%enlarge y limits={value=.1,upper},
%%axis x line*=bottom,
%%axis y line*=left,
%%y axis line style={opacity=0},
%tickwidth=0pt,
%enlarge x limits=true,
xlabel={\# cores},
%ylabel={Effective Bandwidth [GB/s]},
%ylabel={Bandwidth [\GBS]},
%ylabel={Effective BW [GB/s]},
%%%%%%xlabel style={},
%%%%%%ylabel style={},
%%%%%%x tick label style={},
%%%%%%y tick label style={},
%minor x tick num = 7,
xmin=0,
xmax=12,
ymin=0,
ymax=295,
xtick = {0,4,8,12},
ytick = {0,  50, 100, 150, 200, 250},
yticklabels={},
legend pos=south east,
legend style={draw=none},
label style={font=\small},
tick label style={font=\small},
]
\addplot[black, mark=*, mark size=1.5pt, %mark options={fill=white},
thick] table[x expr=\thisrowno{0}, y expr=\thisrowno{2}*0.001, header=true, col sep=comma] {tikz_data/stencil_multicore/u8.txt};
%\addlegendentry{Measured}

\addplot[gray, mark=triangle*, mark size=1.5pt, %mark options={fill=white},
thick] table[x expr=\thisrowno{0}, y expr=\thisrowno{2}*0.001, header=true, col sep=comma] {tikz_data/stencil_multicore/u1.txt};
%\addlegendentry{ECM}

\addplot[red, dashed, very thick] table[x expr=\thisrowno{0}, y expr=\thisrowno{6}, header=true, col sep=comma] {tikz_data/ecm_multicore.dat};
\end{axis}
\node[fill=white] at (0.88,2.6) {(c) \textsc{2d5pt}};
\end{tikzpicture}%

	\centering
	%
	%  \tikzsetnextfilename{#2}%
	\begin{tikzpicture}
  \centering
  \begin{customlegend}[
      legend columns=3,
      legend style={
      draw=none,
      font=\small,
      anchor=north,
      /tikz/every even column/.append style={column sep=0.25cm}},
      legend entries={u=1, u=8, ECM},% phenomenological ECM},
      ]
    \addlegendimage{gray, mark=triangle*, mark size=1.5pt, thick}
    \addlegendimage{black, mark=*, mark size=1.5pt,	thick}
    \addlegendimage{red, dashed, very thick}
  \end{customlegend}
\end{tikzpicture}%

	\vspace{-0.5em}
	\caption{Multicore scaling within one ccNUMA domain for (a) \textsc{triad}, (b) \textsc{sum}, and
          (c) \textsc{2d5pt} kernels, comparing ECM model with measurement. Data
          without unrolling are shown for reference. Note that the read-only memory bandwidth
          was used as a limit for \textsc{sum}. The working set size for \textsc{triad} and \textsc{sum} was set to $4\,\GB$. For \textsc{2d5pt}, the problem size was chosen as $10000^2$ so
          that the layer condition is broken at L1 but fulfilled at L2.} 
	\label{fig:ecm_multicore}
	\vspace{-0.8em}
\end{figure}
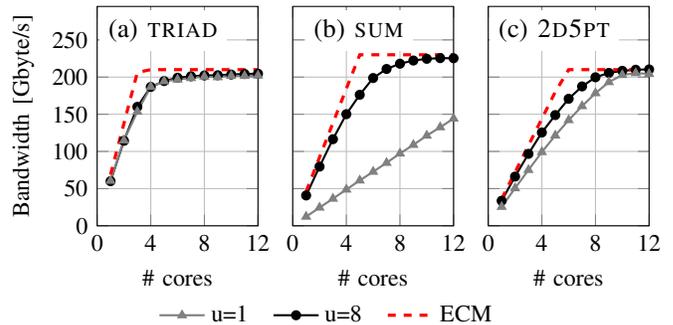

The ``naive'' scaling hypothesis of the ECM model assumes perfect performance scaling
of a loop across the cores of a contention domain (characterized by a shared resource
such as L3 or memory bandwidth) until a bandwidth bottleneck is hit. Figure~\ref{fig:ecm_multicore}
shows a comparison of model and measurement for the \textsc{triad}, \textsc{sum}, and
\textsc{2d5pt} kernels. For \textsc{sum} it is evident that insufficient
MVE (as shown in the ``u=1'' data) is the root cause for non-saturation of the
memory bandwidth due to the long ADD latency. For the stencil kernel, saturation is possible even
without unrolling, but more cores are needed. The ECM model describes the
scaling features qualitatively; the largest deviation occurs around the
saturation point. This is a known effect that can be corrected for by introducing
a latency term~\cite{Hofmann_2018}, which is left for future work.

\section{Case study: Sparse matrix-vector multiplication}\label{sec:spmvm}

%In Figure~\ref{fig:csr_scaling} we saw that \spmv\ similar 
%to the \textsc{sum} reduction benchmark was not capable to saturate the 
%main memory bandwidth. 
%Analysis of the \textsc{sum} benchmark in Section~\ref{subsec:validation}
%further revealed that this is due to the dependency on ADD instruction (see $u=1$ in Figure~\ref{fig:stream_ecm}(b))
%and its associated latency.
%proper unrolling is necessary to hide the pipeline latency. 
The analysis of the \textsc{sum} reduction benchmark revealed that the long ADD latency
prevents bandwidth saturation if proper unrolling with MVE is not employed.
The CRS \spmv\ is similar 
as it requires  horizontal reduction along each row of the matrix (the inner loop).
However, %in contrast to the \textsc{sum} benchmark
this loop is short since the average number of non-zeros per row, $N_\mathrm{nzr}$, is typically in the range of 10--1000
for practical applications.
For the HPCG matrix we have $N_\mathrm{nzr}\approx 27$, so
four 512-bit wide \texttt{fmad}/\texttt{fmla} instructions (eight iterations each) are required,
which takes $4 \times 9\,\cycles$. % (see Table~\ref{table:incore-instructions}). 
The required horizontal reduction (\texttt{faddv}) of the final vector result
adds another 11.5\,\cycles, which leads to a minimum of $47.5\,\cycles$ of in-core
execution time per row ($2.05\,\GFS$).
The memory data traffic for CRS \spmv\ is $\left[27\times(12+8\alpha)+20\right]$\,\bytes\
per row, where $\alpha$ characterizes the efficiency of right-hand side (RHS) access~\cite{KreutzerSELLC14}.
For the regular stencil-like HPCG matrix, $\alpha$ can be assumed to
be close to the optimistic lower limit of $\alpha=1/N_\mathrm{nzr}$, 
resulting in a data traffic of $352$\,\bytes\ per row (measured $363$\,\bytes\ using \likwidperfctr).
%can be estimated using layer condition analysis.
%In this case we have $\alpha=3/N_\mathrm{nzr}$, 
%neglecting the data delay through the cache hierarchy,
This leads to %an upper limit for 
a single-core bandwidth of $352\,\bytes/\mathrm{row}\times f / 47.5\,\cycles/\textrm{row} = 13.3\,\GBS$,
where $f$ is the clock frequency.
Since $12\times 13.3\,\GBS=160\,\GBS$, it is impossible for this code to hit the
bandwidth limit of 210\,\GBS\ per CMG. 
%Even with this simplified light-speed estimate
%the theoretical attainable bandwidth of the code
%is below the measured main memory bandwidth of \afx\ ($48*13.9\,\GBS < 810\,\GBS$). \CAcomm{Should we change to CMG}

In practice, the loop overheads caused by the
extremely short loops and the poor OoO execution
(see Sect.~\ref{subsec:validation}) further decrease the single-core performance.
Contrary to the \textsc{sum} benchmark, unrolling does not fix the problem
due to the extra work at the end of the short loop and loss of 
efficiency when  $N_\mathrm{nzr}$ is not a multiple of the vector length.
%Therefore, a change of data layout is inevitable 
%to saturate the high bandwidth of the machine. 
%For almost all other  standard server CPUs in the market today, this is not the case since even with a weak
%single-core performance they are able to saturate their relatively low memory bandwidth.
Hence, we choose the \sellcs\ sparse matrix format~\cite{KreutzerSELLC14}, which is
a portable, SIMD-friendly data format for CPUs, GPUs, and vector machines.

\sellcs\ stores chunks of $C$ consecutive rows (zero-padded to the longest row)
in column-major format.
%rather than the usual row-major format used in CRS. \CAcomm{Is it OK, since it is not actualy col major, it is nnzr major?}
The parameter $C$ is tunable; for efficiency, it should be a multiple of VL as well as large enough to allow for sufficient unrolling.
%since usually the sparse matrices have
%large number of rows a sufficiently big $C$ can be chosen  to support long VL and unrolling.
A further benefit of the format is the lack of an expensive horizontal
add operation (\texttt{faddv}). % and avoids the reduction of vector 
%efficiency as $C$ is always chosen as a multiple of VL.
The only drawback is that the zero padding can reduce the efficiency if
rows have very different lengths. 
%each chunk of rows has to be padded along the 
%column direction to the longest row, which increases fill-in with zero elements reducing the efficiency ($\beta$).
To mitigate this effect, rows are first sorted by descending length within a sorting window ($\sigma$) to
reduce the padding. With proper selection of $C$ and $\sigma$, the padding can
be made negligible in most cases.
For the matrices considered in this work it was never larger than 5\%.

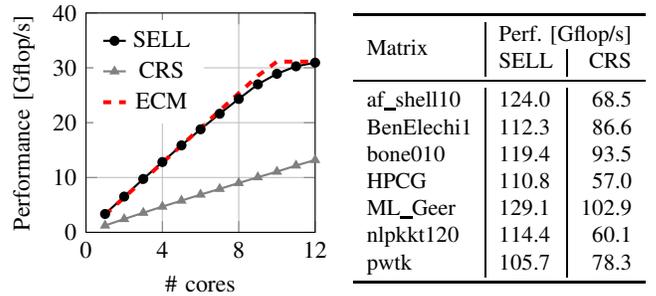
\begin{figure}[tb]
	\centering
	\begin{minipage}{0.55\linewidth}
		\centering
		%
	%  \tikzsetnextfilename{#2}%
	\pgfplotsset{
	compat=1.11,
	legend image code/.code={
		\draw[mark repeat=2,mark phase=2]
		plot coordinates {
			(0cm,0cm)
			(0.15cm,0cm)        %% default is (0.3cm,0cm)
			(0.3cm,0cm)         %% default is (0.6cm,0cm)
		};%
	}
}

\begin{tikzpicture}%[y=.2cm, x=.7cm]%,font=\sffamily]
\centering
\begin{axis}[
%%scale only axis,
height=4.5cm,
width=0.95\linewidth,
grid=both,
tick align=inside,
%axis y line*=left,
%%enlarge y limits={value=.1,upper},
%%axis x line*=bottom,
%%axis y line*=left,
%%y axis line style={opacity=0},
%tickwidth=0pt,
%enlarge x limits=true,
xlabel={\# cores},
ylabel={Performance [\GFS]},
%ylabel={Effective BW [GB/s]},
%%%%%%xlabel style={},
%%%%%%ylabel style={},
%%%%%%x tick label style={},
%%%%%%y tick label style={},
%minor x tick num = 7,
xmin=0,
xmax=12,
ymin=0,
ymax=40,
xtick = {0,4,8,12},
legend pos=north west,
legend style={draw=none, font=\small},
label style={font=\small},
tick label style={font=\small},
]
\addplot[black, mark=*, mark size=1.5pt, %mark options={fill=white},
thick] table[x expr=\thisrowno{0}, y expr=\thisrowno{2}, header=true, col sep=comma] {tikz_data/scaling_hpcg_sell32.dat};
\addlegendentry{SELL}

\addplot[gray, mark=triangle*, mark size=1.5pt, %mark options={fill=white},
thick] table[x expr=\thisrowno{0}, y expr=\thisrowno{1}, header=true, col sep=comma] {tikz_data/scaling_hpcg_csr_u1.dat};
\addlegendentry{CRS}

\addplot[red, dashed, thick, line width=1.5 pt] table[x expr=\thisrowno{0}, y expr=\thisrowno{7}, header=true, col sep=comma] {tikz_data/ecm_hpcg_sell32.dat};
\addlegendentry{ECM}

\end{axis}
%\begin{axis}[
%%%scale only axis,
%height=4.5cm,
%width=0.95\linewidth,
%%grid=both,
%tick align=inside,
%axis y line*=right,
%%%enlarge y limits={value=.1,upper},
%%%axis x line*=bottom,
%%%axis y line*=left,
%%%y axis line style={opacity=0},
%%tickwidth=0pt,
%%enlarge x limits=true,
%xlabel={\# cores},
%%ylabel={Effective Bandwidth [GB/s]},
%ylabel={Bandwidth [\GBS]},
%%ylabel={Effective BW [GB/s]},
%%%%%%%xlabel style={},
%%%%%%%ylabel style={},
%%%%%%%x tick label style={},
%%%%%%%y tick label style={},
%%minor x tick num = 7,
%xmin=0,
%xmax=12,
%ymin=0,
%ymax=260.8955223880597,
%xtick = {0,4,8,12},
%ytick = {0,  50, 100, 150, 200},
%%ytick = {0,20,40,60,80,100,120,140},
%legend pos=south east,
%legend style={draw=none},
%]
%\addplot[black, mark=*, mark size=1.5pt, %mark options={fill=white},
%thick] table[x expr=\thisrowno{0}, y expr=\thisrowno{2}*0.5*(12+28/26.8), header=true, col sep=comma] {tikz_data/scaling_hpcg_sell32.dat};
%%\addlegendentry{Measured}
%
%\addplot[gray, mark=triangle*, mark size=1.5pt, %mark options={fill=white},
%thick] table[x expr=\thisrowno{0}, y expr=\thisrowno{1}*0.5*(12+28/26.8), header=true, col sep=comma] {tikz_data/scaling_hpcg_csr_u1.dat};
%
%\addplot[red, dashed, thick, line width=1.5 pt] table[x expr=\thisrowno{0}, y expr=\thisrowno{7}*0.5*(12+28/26.8), header=true, col sep=comma] {tikz_data/ecm_hpcg_sell32.dat};
%%\addlegendentry{ECM}
%
%\end{axis}
%%\node[fill=white] at (1.24,2.5) {(a) intra-CMG};
\end{tikzpicture}
%\hspace{-0.7em}
%
	
	\end{minipage}%
	\begin{minipage}{0.45\linewidth}
		\centering
		\resizebox{\linewidth}{!}{%
		  \begin{tabular}{l| l| r}
		  	\toprule

      \multirow{2}{*}{Matrix}   & \multicolumn{2}{c}{Perf. [\GFS]}\\
%      	\cline{2-3}
	 &  {SELL} & {CRS} \\
	\midrule
	{af\_shell10} &   {124.0} & {68.5}\\
	{BenElechi1} &   {112.3} & {86.6}\\
	{bone010} &  {119.4} & {93.5}\\
	{HPCG} &  {110.8} & {57.0}\\
	{ML\_Geer} &  {129.1} & {102.9}\\
	{nlpkkt120} &   {114.4} & {60.1}\\
	{pwtk} &  105.7 & {78.3}\\
			\bottomrule
		\end{tabular}
		} % resizebox
%	\end{table}
	%	\caption{table caption goes  here}\label{tab: table-label}
	\end{minipage}
\caption{(Left) Scaling performance and ECM prediction on one CMG for \spmv\ in \sellcs\ 
	and CRS format with the HPCG matrix. (Right) Full-node \spmv\ 
	performance of different matrices in \GFS\ in both formats. $C$ was chosen as 32 
	and $\sigma$ was tuned between 1 and 1024. Reverse Cuthill-McKee reordering \cite{RCM}
        was done if it improved the
        performance.
}\label{fig:sell_c_perf}
\vspace{-0.8em}
\end{figure}

On \afx, $C = 32$ enables four-way unrolling on top of SVE, which
should reduce the ADD latency by a factor of four (i.e., to 2.25\,\cycles).
According to the ECM model, the  data transfer through the memory hierarchy
should then become the bottleneck.  Much of the time here is lost
in the LD to the index array and the subsequent gather instruction from the L1 cache to the registers
% corresponding to the right hand side vector that costs approximately
(5.5\,\cycles\ according to Table~\ref{table:incore-instructions}).
Loading the matrix values costs an extra 0.5\,\cycles, so we require at least
$(5.5+0.5)\times27/8\,\cycles=20.3\,\cycles$ (at $\mathrm{VL}=8$)
per row of the HPCG matrix for the L1 to register transfers. 
From L2 and memory, as shown above, at least $352\,\bytes$ are required per row,
which translates to $352/64\,\cycles=5.5\,\cycles$  and $352/117\,\cycles=3\,\cycles$ respectively
(see Table~\ref{table:testbed} for bandwidths).
%Finally, as shown above, $368\,\bytes$ must come from memory,
%requiring $368/117\,\cycles=3.1\,\cycles$. 
Since the transfers are primarily reads, the contributions have to be summed up as shown
in the overlap hypothesis in Sect.~\ref{subsec:overlap_hypo}. 
This results in a total of almost $28.8\,\cycles$ per row,
i.e., $3.4\,\GFS$ (equivalent memory bandwidth $22\,\GBS$). 
The measured performance is $3.3\,\GFS$ on a single core.
Hence, the code can saturate on a CMG
as shown in Fig.~\ref{fig:sell_c_perf} (left), topping out at
$31\,\GFS$ (i.e., $202\,\GBS$).
In contrast to other contemporary CPUs, saturation requires almost all cores
of a CMG; any loss of efficiency on the sequential code due to
extra traffic or latency will directly impact the total performance. 
The scaling across ccNUMA domains (using parallel first touch) is nearly
perfect, leading to a final performance of $110.8\,\GFS$. Based on our
previous Roof{}line analysis of the HPCG algorithm~\cite{Understanding2020}, we expect a full HPCG
benchmark performance of close to 100\,\GFS\ on \afx.

In Fig.~\ref{fig:sell_c_perf} (right) we show the full-node (48 cores) performance
of \sellcs\ 
in comparison to CRS on a collection of matrices from the
SuiteSparse Matrix Collection~\cite{UOF}.
All matrices saturated the main memory bandwidth with \sellcs,
and we see an improvement of $1.5 \times$ compared to CRS on average.
The \spmv\ performance attained with \sellcs\ is on par with
NVIDIA V100 GPU~\cite{Tsai20V100} and NEC SX-Aurora Tsubasa
vector accelerators~\cite{GomezTsubasa}.

%\begin{table}[!tb]
%	\centering
%	\caption{Performance of \spmv}
%	\label{table:spmv_perf}
	%	\resizebox{\textwidth}{!}{%
%	\begin{tabular}{l l}
%		\toprule
%{Matrix} &  {\GFS} \\
%{af\_shell10} &   {124.0} \\
%{BenElechi1} &   {112.3} \\
%{bone010} &  {119.4} \\
%{HPCG} &  {110.8} \\
%{ML\_Geer} &  {129.1} \\
%{nlpkkt120} &   {114.4} \\
%{pwtk} &  105.7 \\
%		\bottomrule
%	\end{tabular}
	%	} % resizebox
%\end{table}

\section{Conclusion}\label{sec:summary}

\subsection{Summary and outlook}
Via an analysis of in-core features and data transfers, we have established an ECM machine model
for the \afx\ CPU in the Fujitsu FX700 system and applied it to simple streaming kernels and sparse matrix-vector
multiplication using the HPCG matrix.
For in-memory data sets,
the single-core ECM model was shown to be accurate within a maximum error of 20\%.
The memory hierarchy turned out to be
partially overlapping, allowing for a substantial single-core memory bandwidth with optimized code.
Long floating-point instruction latencies and
limited out-of-order execution capabilities were identified as the main culprits
of poor performance and lack of bandwidth saturation. With the current gcc compiler,
vector intrinsics and manual unrolling are often required to achieve
high performance.
For \spmv, the \sellcs\ matrix storage was shown to achieve superior performance
and memory-bandwidth saturation, though requiring almost all cores on the ccNUMA domain.
This means that \spmv\ performance will be very sensitive to load imbalance and
inefficiencies in data accesses. 

In future work we will conduct a more comprehensive analysis of \spmv\ and a comparison
with other high-end devices like the Nvidia A100 GPU and the NEC SX-Aurora Tsubasa.
We will also investigate advanced \afx\  features such as the sector cache and cache line
zero instructions. In addition, we will refine the ECM model with a more accurate
saturation model using an additional latency term.
We will also apply the refined model to the HPCG benchmark and to
lattice QCD applications~\cite{Meyer2020}. 

\subsection{Related work}
%Since the A64FX is publicly available only for a short period at the
%time of writing, only two works related to this paper are known to the
%authors.  J. Dongarra~\cite{Dongarra20Fugaku} gives a broad overview
%about Fugaku, its software stack and the performance of basic
%benchmarks on the system.  Jackson et al.~\cite{jackson2020A64FX}
%investigate the porting and performance of several mini-apps and
%standard benchmarks on Fugaku compared to other modern Intel and Arm
%systems and evaluate the out-of-the-box efficiency for those codes on
%the A64FX.

Since the \afx\ CPU has only been available for a short time,
the amount of performance-centric research is limited. \textsc{Dongarra}~\cite{Dongarra20Fugaku}
reports on basic architectural features, HPC benchmarks (HPL, HPCG,
HPL-AI), and the software environment of the Fugaku system.
\textsc{Jackson} et al.~\cite{jackson2020investigating} 
investigate some full applications and proxy apps in comparison to
Intel and other Arm-based systems but do not use performance models
for analysis.

%\begin{itemize}
%	\item Tell about high latency of ADD for \spmv
%	\item OoO
%	\item Tell GCC compiler is not able to do MVE unrolling and such kind of optimisations
%\end{itemize}
%\begin{itemize}
%\item More complete coverage of \spmv, comparison w/ other CPUs
%\item FX1000
%\end{itemize}

%\section*{Acknowledgment}
%
%Do we need this? Any funding to be mentioned?

\bibliographystyle{IEEEtran}
\bibliography{pub}
\vspace{12pt}
\end{document}